# Crossing point phenomena (T* = 2.7 K) in specific heat curves of superconducting ferromagnets $RuSr_2Gd_{1.4}Ce_{0.6}Cu_2O_{10-\delta}$

Anuj Kumar[1,2,*], R. P. Tandon[2], Jianli Wang[3], Rong Zeng[3] and V. P. S. Awana[1,†]

[1]*Quantum Phenomena Application Division, National Physical Laboratory (CSIR)*
*Dr. K. S. Krishnan Road, New Delhi-110012, India*

[2]*Department of Physics and Astrophysics, University of Delhi, North Campus*
*New Delhi-110007, India*

[3]*Institute for Superconducting and Electronic Materials, University of Wollongong, Northfields Avenne, Wollongong New South Wales 2522, Australia*

Crossing point phenomena are one of the interesting and still puzzling effects in strongly correlated electron systems. We have synthesized $RuSr_2Gd_{1.4}Ce_{0.6}Cu_2O_{10-\delta}$ (GdRu-1222) magneto-superconductor through standard solid state reaction route and measured its magnetic, transport and thermal properties. We also synthesized $RuSr_2Eu_{1.4}Ce_{0.6}Cu_2O_{10-\delta}$ (EuRu-1222) then measured its heat capacity in zero magnetic fields for reference. The studied compounds crystallized in tetragonal structure with space group *I4/mmm*. GdRu-1222 is a reported magneto-superconductor with Ru spins magnetic ordering at temperature around 110 K and superconductivity in Cu-$O_2$ planes below around 40 K. To explore the crossing point phenomena, the specific heat [$C_p(T)$] was investigated in temperature range 1.9-250 K, under magnetic field of up to 70 kOe. Unfortunately though no magnetic and superconducting transitions are observed in specific heat, a Schottky type anomaly is observed at low temperatures below 20 K. This low temperature Schottky type anomaly can be attributed to splitting of the ground state spectroscopic term $^8S_{7/2}$ of paramagnetic $Gd^{3+}$ ions by both internal and external magnetic fields. It was also observed that $C_p(T)$ being measured for different values of magnetic field, possesses the same crossing point (T* = 2.7 K), up to the applied magnetic field 70 kOe. A quantitative explanation of this phenomenon, based on its shape and temperature dependence of the associated generalized heat capacity ($C_p$), is presented. This effect supports the crossing point phenomena, which is supposed to be inherent for strongly correlated systems.

# I. Introduction

Discovery of the co-existence of superconductivity and magnetic order in the hybrid rutheno-cuperate systems $RuSr_2(Eu,Gd,Sm)_{1.5}Ce_{0.5}Cu_2O_{10-\delta}$ (RERu-1222) and $RuSr_2(Eu,Gd,Sm)Cu_2O_{8-\delta}$ (RERu-1212) is interesting because the magnetic ordering temperature is much higher than the superconducting transition temperature [1-5]. However, despite the extensive research on these materials, some questions yet remain unanswered. For instance, the oxygen non-stoichiometry, carrier concentration and valence state of Ru are also still inconclusive. The first simultaneous observation of superconductivity and the Ru ion magnetic order in a rutheno-cuprate was published in 1997 by *I. Felner* et al. [1]. These systems show ferromagnetic ordering at a higher Curie temperature of around 100-135 K and superconductivity at a lower critical temperature of about 15-40 K. The superconductivity is associated with the $Cu-O_2$ planes, while the magnetic order originates from the $Ru-O_2$ planes. These $Ru-O_2$ planes are not only responsible for weak ferromagnetism but also provide charge carriers to the superconducting $Cu-O_2$ planes.

Very recently, a crossing point phenomenon is seen due to ground state spectroscopic term splitting of $Gd^{3+}$ ion in heat capacity of GdRu-1222 [6]. The crossing point phenomenon is one of the interesting and still puzzling effects in strongly correlated electron systems [7-9]. A typical example of this effect is the temperature dependence of specific heat curves $C_p$ $(T, X)$ observed at different values of a thermodynamic variable X (such as magnetic field $H$ or pressure $P$). This type of effect was found not only for thermodynamic but for dynamical quantities i.e., frequency dependent optical conductivity etc. Generally, this familiar effect is named as *isobestic point* [10]. The curves cross at one particular temperature known as crossing point temperature T*. In particular, the crossing point in $C_p$ $(T, H)$ curves was found in the heavy-fermion compounds like $CeCu_{5.5}Au_{0.5}$ [11], semi-metallic $Eu_{0.5}Sr_{0.5}As_3$ [12], superconducting cuprate $GdBa_2Cu_3O_{7-\delta}$ [13] and manganite $NdMnO_3$ [14]. The known theoretical considerations [7-10] are based on rather different approaches. Available relevant experimental data are inconclusive, therefore more experimental data in support this phenomenon are warranted for different strongly correlated electron systems. Till now the general reasons and conditions for occurrence of *isobestic points* are not clear. In this report, the crossing point effect is shown in $C_p$ $(T, H)$ curves of polycrystalline known superconducting ferromagnets $RuSr_2Gd_{1.4}Ce_{0.6}Cu_2O_{10-\delta}$ (GdRu-1222). An interesting phenomenon associated with the crossing point is the Schottky anomaly, where the specific heat capacity of a solid at low temperature makes a "*bump*". It is called anomalous because the heat capacity usually increases with temperature or remains constant [15].

We present and discuss the crossing point phenomenon in GdRu-1222 and study specific features of the magnetic state of this compound. To compare the magnetic effects of different rare-earths, we also measured low temperature heat capacity of the sample $RuSr_2Eu_{1.4}Ce_{0.6}Cu_2O_{10-\delta}$ (EuRu-1222) and compared with GdRu-1222.

## II. Experimental Details

Polycrystalline bulk samples of $RuSr_2Gd_{1.4}Ce_{0.6}Cu_2O_{10-\delta}$ (GdRu-1222) and $RuSr_2Eu_{1.4}Ce_{0.6}Cu_2O_{10-\delta}$ (EuRu-1222) are synthesized through standard solid state reaction route from stoichiometric powders of 99.99% purity $RuO_2$, $SrCO_3$, $Gd_2O_3$, $Eu_2O_3$, $CeO_2$ and $CuO$. These mixtures were ground together in an agate mortar and calcined in air atmosphere at $1020^{o}C$, $1040^{o}C$ and $1060^{o}C$ each for 24 hrs with intermediate grindings. The pressed bar shaped pellets were annealed in Oxygen atmosphere at $850^{o}C$, $650^{o}C$ and $450^{o}C$ each for 24 hrs, and subsequently cooled down slowly over a span of 12 hrs to the room temperature. X-ray diffraction (*XRD*) of studied GdRu-1222 was done at room temperature in the scattering angular ($2\theta$) range of $20^{o}$-$80^{o}$ in equal steps of $0.02^{o}$ using *Rigaku* diffractometer with $Cu\ K_\alpha$ ($\lambda = 1.54$ Å) radiation. Detailed Rietveld analysis was performed using the *FullProf* program. *DC* magnetization was performed on Physical Property Measurements System (PPMS-14T, Quantum Design-USA) as a function of temperature. Specific heat measurements over the temperature range of 1.9 K to 250 K for EuRu-1222 in zero fields and for GdRu-1222 in various fields of up to 70 kOe were also performed also on PPMS. Resistivity measurements were performed on a Closed Cycle Refrigerator (CCR) over a temperature range 10-300 K using the simple four probe geometry.

## III. Results and Discussion

The quality of sample in case of rutheno-cuprates is very important for any meaningful discussion, since minute impurities of $SrRuO_3$ and $Sr_2GdRuO_6$ ($211O_6$) phases tend to form readily in the matrix. Any small impurity of these foreign phases may complicate the resultant magnetization and other physical properties. As seen from Figure 1, characteristic peaks corresponding to $SrRuO_3$ and $Sr_2GdRuO_6$ phases are not observed and rather a phase pure GdRu-1222 is seen within the resolution limit of XRD. Observed and fitted X-ray pattern for the compound is shown in Figure 1. The structural analysis was performed using the Rietveld refinement analysis by using the *FullProf* Program. The Rietveld analysis confirms a tetragonal structure single phase formation in *I4/mmm* space group. All the Reitveld refined structural parameters (lattice parameters, site occupancy and atomic coordinates) of compound are shown in Table I.

The temperature behavior of magnetization, *M-T* shows important features of the complicated magnetic state of this compound. Main panel of Figure 2 depicts the temperature dependence of zero-field-cooled (*ZFC*) and field-cooled (*FC*) *DC* magnetization plots for the studied $RuSr_2Gd_{1.4}Ce_{0.6}Cu_2O_{10-\delta}$ (GdRu-1222) magneto-superconductor, being measured at 20 Oe applied field. The sharp rise of both *ZFC* and *FC* curve at (Curie temperature) $T_C = 120$ K shows a paramagnetic (*PM*) to ferromagnetic (*FM*) transition. The large difference between the *ZFC* and *FC* curves exhibits spin-glass/cluster glass state in the system. The system enters into a new phase called glassy phase [16] just below the Curie temperature ($T_C$). The *ZFC* curve shows a peak at around $T_f = 80$ K, just below the temperature, where *ZFC* and *FC* curve separates. In *ZFC* curve the system undergoes a superconducting transition at $T_c = 27$ K. Inset of Figure 2 shows the temperature dependence of Resistance of studied GdRu-1222 from temperature range 10-300 K. As we move from high temperature to low temperature the resistance increases slightly exhibiting the insulator like behavior. But just below the freezing temperature ($T_f = 80$ K), the resistance increases very fast and finally drops sharply at SC transition at $T_c = 27$ K.

Figure 3 reveals specific heat at low temperature range (below 40 K) of studied GdRu-1222 and EuRu-1222 in zero magnetic fields. Interestingly no magnetic (~150 K) or superconducting transitions (~30 K) are observed in both the samples, see inset Figure 3. Rather, an upturn in heat capacity $C_p$ (T) below 20 K (Schottky-type anomaly) is observed for GdRu-1222 sample. This low-temperature Schottky-type anomaly below 20 K can be attributed due to splitting of the ground state term $^8S_{7/2}$ of paramagnetic ions by internal and external magnetic fields. According to the Kramer's Theorem, the degenerate ground state term should split into four doublets in tetragonal symmetry. Internal molecular fields can arise in rutheno-cuprate from both the Gd and Ru sub-lattice [17] and can co-exist with superconductivity. Even though a direct Gd-Gd exchange interaction is unlikely, these ions can be magnetically polarized by the

*4d-4f* interaction, as discussed in more details in ref. 18. Generally, the Schottky term in the specific heat for the compounds with $Gd^{3+}$ ions should be attributed to the splitting of all four doublets, although actually only some of them make the dominant contribution to the effect. The ground state term for Eu ions is $^7F_0$, hence no splitting takes place on the application of magnetic field. The EuRu-1222 sample shows smooth $C_p(T)$ dependences, the curve is of the Debye type without any low temperature magnetic anomaly. Also, it should be noted that no magnetic transition (Curie temperature $T_C$ = 100-120 K) was observed in heat capacity for both GdRu-1222 and EuRu-1222. This can be explained due to the absence of long range magnetic order in these systems [19]. There may be magnetic in-homogeneities as well at the transition point.

Figure 4 shows the specific heat curves for GdRu-1222 sample observed at different value of applied magnetic field. All heat capacity curves cross at the same temperature point (known as crossing temperature) $T^*$ = 2.7 K with the specific heat value $C_p^*$ = 7.04 J/mole-K. This happens only for the GdRu-1222 and not for EuRu-1222 sample. The crossing point effect is considered as some special type of universality for strongly correlated electron systems, but still no defined theory or mechanism for this phenomenon. Only some general reasons and preliminary thoughts are provided for it occurrence. It is mentioned in [10], that the crossing point appears in a system with superposition of two (or more) components, like that in the known Gorter-Casimir two fluid model of superconductivity. In consideration of the crossing effect in $C_p(T, H)$ curves, it is important to know the motive force for strong magnetic field dependence of specific heat. For GdRu-1222, the motive force is associated with splitting of the ground state term $^8S_{7/2}$ of paramagnetic $Gd^{3+}$ ions by internal and external magnetic fields [18].

The Schottky anomaly in the $C_p(T, H)$ curves is itself a background for the crossing effect found in this study. There is a thermodynamics approach [6], which can be helpful to understand this phenomenon. It can be suggested that in the low temperature range, where the crossing point phenomena takes place, the magnetic contribution to the specific heat is dominant. The expression for the specific heat at constant H is $C_H = T(dS/dT)_H$. Only if $T^*$ is independent of $H$, will all $C_H(T, H)$ curves intersect in one point demonstrating a true crossing effect (see Figure 4).

## IV. Conclusions

We synthesized polycrystalline magneto-superconducting samples GdRu-1222 and EuRu-1222 by solid state reaction route and measured their structural, magnetic and heat capacity. The crossing point phenomenon in GdRu-1222 is observed and discussed. To compare the paramagnetic effects of different rare-earth compounds, we also measured magnetization and low temperature heat capacity of EuRu-1222. The low temperature Schottky type anomaly is observed only in GdRu-1222 compound and not in EuRu-1222. This is due to the splitting of the ground state term $^8S_{7/2}$ of paramagnetic ions $Gd^{3+}$ by applied internal and external magnetic fields. We hope that results of this study will help to adequately model the crossing point phenomenon in $C_p$ *(T, H)* curves for magnetic systems, having a transition from classical to quantum behavior in $C_p$ *(T, H)* with decreasing temperature.

## Acknowledgements

The authors would like to thank Prof. R. C. Budhani (Director, NPL) for his constant support and encouragement during the work. One of the authors Mr. Anuj Kumar would like to thanks Council of Scientific and Industrial Research (CSIR), for providing financial support through Senior Research Fellowship (SRF) to pursue his Ph.D. This work is also financially supported by Department of Science and Technology (DST-SERC), New Delhi funded project on Investigation of pure and substituted Rutheno-cuprate magneto-superconductors in bulk and thin film form at low temperature and high magnetic field.

# Figure Captions

**Figure 1** Observed (*solid circles*) and calculated (*solid lines*) XRD patterns of $RuSr_2Gd_{1.4}Ce_{0.6}Cu_2O_{10-\delta}$ compound at room temperature. Solid lines at the bottom are the difference between the observed and calculated patterns. Vertical lines at the bottom show the position of allowed Bragg peaks.

**Figure 2** Zero-Field-Cooled (*ZFC*) and Field-Cooled (*FC*) *DC* magnetization plot for $RuSr_2Gd_{1.4}Ce_{0.6}Cu_2O_{10-\delta}$ measured in the applied magnetic field, H = 20 Oe. Inset shows the R vs. T plot for $O_2$-annealed GdRu-1222.

**Figure 3** Low temperature behavior of total specific heat at zero magnetic fields for samples GdRu-1222 ($RuSr_2Gd_{1.4}Ce_{0.6}Cu_2O_{10-\delta}$) and EuRu-1222 ($RuSr_2Eu_{1.4}Ce_{0.6}Cu_2O_{10-\delta}$). Inset shows heat capacity of same for temperature range 1.9-250 K.

**Figure 4** Low temperature dependences of total specific heat ($C_p$), taken at different values of applied magnetic field for $RuSr_2Gd_{1.4}Ce_{0.6}Cu_2O_{10-\delta}$. These $C_p$ (T) curves cross each other at the same temperature T* = 2.7 K, reveling the crossing point phenomena.

**Table I.** Atomic coordinates and site occupancy for studied $RuSr_2Gd_{1.4}Ce_{0.6}Cu_2O_{10-\delta}$

Space group: *I4/mmm*, Lattice parameters; *a* = 3.8352 (4) Å, *c* = 28.5728 (3) Å, $\chi^2$ = 3.76

| Atom | Site | x (Å) | y (Å) | z (Å) |
|------|------|-------|-------|-------|
| **Ru** | 2b | 0.0000 | 0.0000 | 0.0000 |
| **Sr** | 2h | 0.0000 | 0.0000 | 0.4205 (4) |
| **Gd/Ce** | 1c | 0.0000 | 0.0000 | 0.2942 (6) |
| **Cu** | 4e | 0.0000 | 0.0000 | 0.1455 (3) |
| **O(1)** | 8j | 0.6152 (3) | 0.5000 | 0.0000 |
| **O(2)** | 4e | 0.0000 | 0.0000 | 0.0741 (2) |
| **O(3)** | 8g | 0.0000 | 0.5000 | 0.1461 (3) |
| **O(4)** | 4d | 0.0000 | 0.5000 | 0.2500 |

**Figure 1**

**Figure 2**

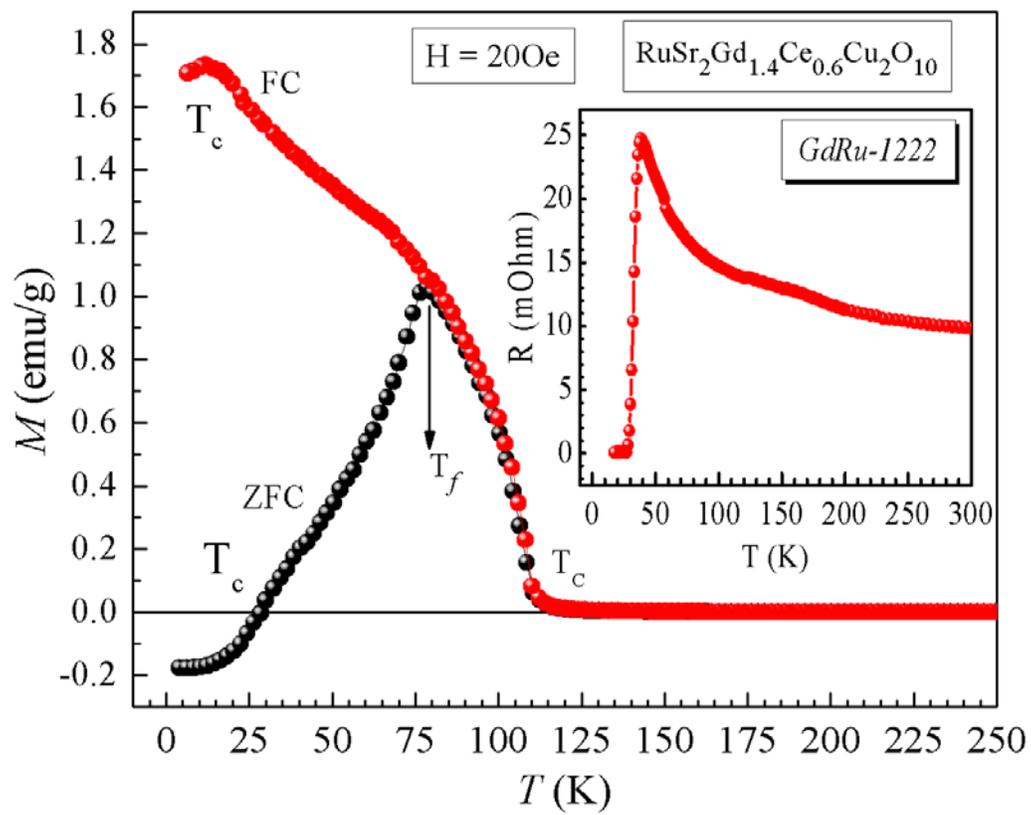

**Figure 3**

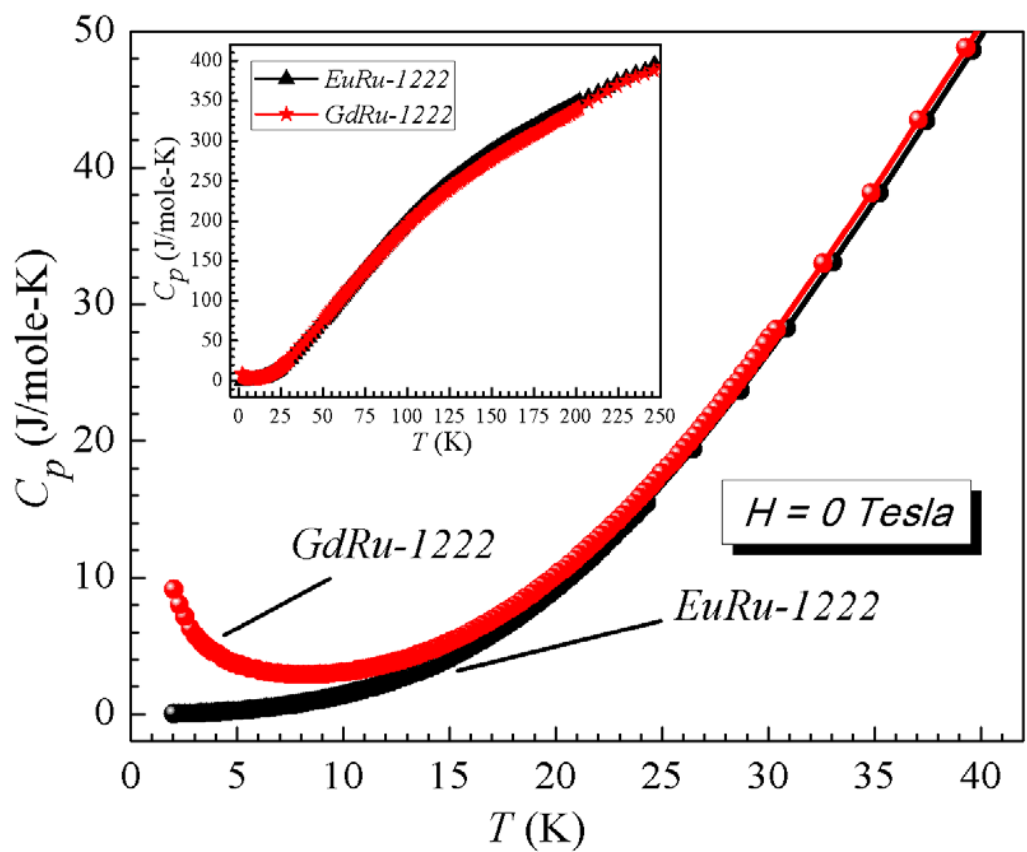

**Figure 4**

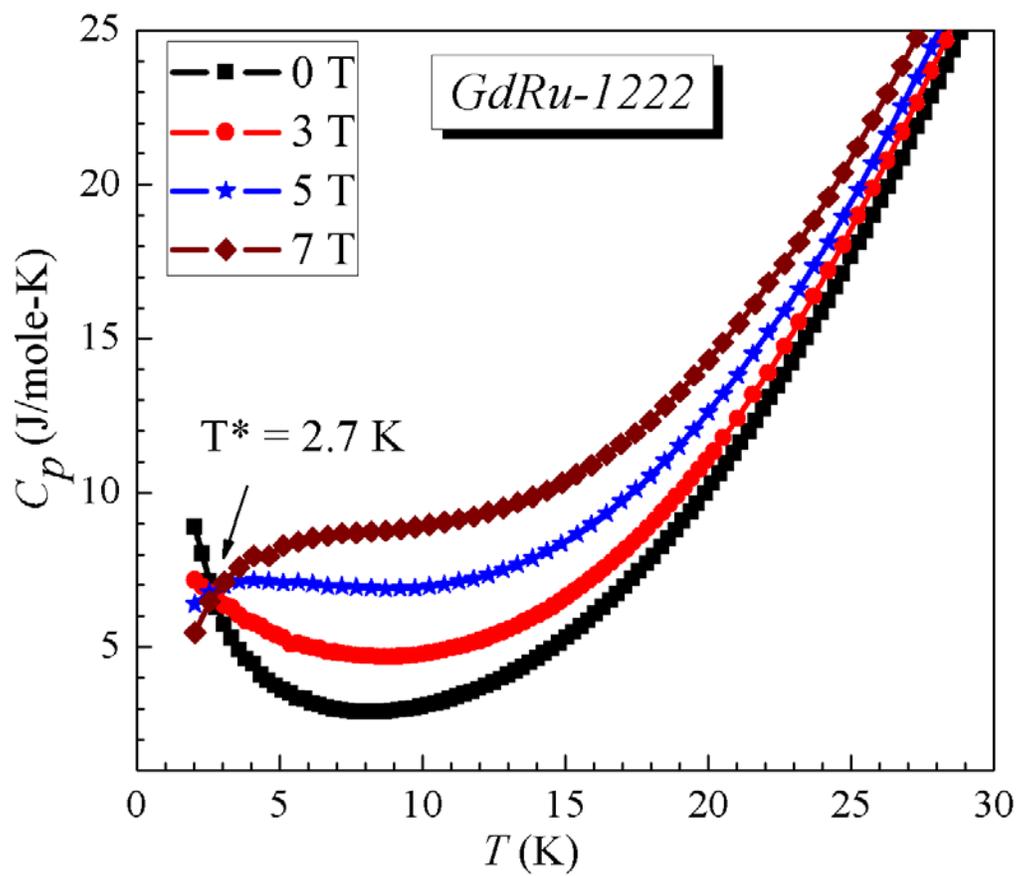